\documentclass[journal]{IEEEtran}

\usepackage{graphicx}
\usepackage{stackengine}

\ifCLASSINFOpdf
\else
\fi
\usepackage{amsmath}
\usepackage{amsfonts} 
\usepackage{hyperref}
\usepackage{subfigure}
\usepackage{algorithm}
\usepackage{algpseudocode}
\usepackage{booktabs}
\ifCLASSOPTIONcompsoc
\else
\fi
\usepackage{caption}
\usepackage[labelformat=simple]{subcaption}
 
\usepackage{dsfont}
\usepackage[table,xcdraw]{xcolor}
\usepackage{amssymb}
\usepackage{amsthm}
\newtheorem{definition}{Definition}
\newtheorem{lemma}{Lemma}

\usepackage{textcase}
\makeatletter

\renewcommand{\section}{\@startsection{section}{1}{\z@}{-3.5ex \@plus -1ex \@minus -.2ex}{2.3ex \@plus.2ex}{\normalfont\normalsize\centering\bfseries\MakeUppercase}}
\renewcommand{\subsection}{\@startsection{subsection}{2}{\z@}{-3.25ex\@plus -1ex \@minus -.2ex}{1.5ex \@plus .2ex}{\normalfont\normalsize\MakeUppercase}}
\makeatother
\usepackage{paralist}

\begin{document}

\title{Delay Tolerant Networking to Extend Connectivity in Rural Areas Using Public Transport Systems: Design And Analysis}
\author{Salah~Abdeljabar,~\IEEEmembership{Graduate~Student~Member,~IEEE}, Marco~Zennaro,~\IEEEmembership{Senior~Member,~IEEE}, and Mohamed-Slim~Alouini,~\IEEEmembership{Fellow,~IEEE}
\thanks{Salah Abdeljabar and Mohamed-Slim Alouini are with the Computer, Electrical and Mathematical Science and Engineering Division, King Abdullah University of Science and Technology (KAUST), Thuwal 23955-6900, Saudi Arabia (e-mail: salah.abdeljabar@kaust.edu.sa, slim.alouini@kaust.edu.sa). Marco Zennaro is with the Abdus Salam International Centre for Theoretical Physics, Trieste, Italy (e-mail: mzennaro@ictp.it).}% <-this % stops a space
}

\markboth{IEEE Internet of Things Journal}%
{Shell \MakeLowercase{\textit{et al.}}: Bare Demo of IEEEtran.cls for IEEE Journals}

\maketitle
\begin{abstract}
In today's digital age, access to the Internet is essential, yet a significant digital divide exists, particularly in rural areas of developing nations. This paper presents a Delay Tolerant Networking (DTN) framework that utilizes informal public transportation systems, such as minibus taxis, as mobile data mules to enhance connectivity in these underserved regions. We develop a probabilistic model to capture the randomness in vehicle mobility, including travel times and contact durations at bus stops. Key performance metrics are analyzed, including average data transmission rate and Peak Age of Information (PAoI), to assess the effectiveness of the proposed system. An analytical approximation for the Mean PAoI (MPAoI) is derived and validated through simulations. Case studies from real-world datasets in Nouakchott, Accra, and Addis Ababa demonstrate the practical applicability and scalability of our framework. The findings indicate that leveraging existing transportation networks can significantly bridge the digital divide by providing reliable internet-like connectivity to remote areas.
\end{abstract}


\IEEEpeerreviewmaketitle
 
%%%%%%%%%%%%%%%%%%%%%%%%%%%%%%%%%%%%%%%%%%%%%
%%%%%%%%%%%%%%%%%%%%%%%%%%%%%%%%%%%%%%%%%%%%%
%%%%%%%%%%%%%%%%%%%%%%%%%%%%%%%%%%%%%%%%%%%%%
\section{Introduction}
%%%%%%%%%%%%%%%%%%%%%%%%%%%%%%%%%%%%%%%%%%%%%
%%%%%%%%%%%%%%%%%%%%%%%%%%%%%%%%%%%%%%%%%%%%%
%%%%%%%%%%%%%%%%%%%%%%%%%%%%%%%%%%%%%%%%%%%%%

%%%%%%%%%%%%%%%%%%%%%%%%%%%%%%%%%%%%%%%%%%%%%
%%%%%%%%%%%%%%%%%%%%%%%%%%%%%%%%%%%%%%%%%%%%%
\subsection{Motivation}
%%%%%%%%%%%%%%%%%%%%%%%%%%%%%%%%%%%%%%%%%%%%%
%%%%%%%%%%%%%%%%%%%%%%%%%%%%%%%%%%%%%%%%%%%%%
\par
In today’s digital age, Internet access is widely recognized as essential for economic and social progress. The United Nations (UN) emphasizes this importance in their Common Agenda, declaring universal Internet access by 2030 as a fundamental human right~\cite{Secretary-General_2021}. Despite rapid advancements in connectivity solutions, a significant disparity in access persists between developing and developed nations. According to the latest reports from the International Telecommunication Union (ITU), approximately 2.9 billion people, which is nearly one-third of the global population, still lack Internet access, predominantly in rural areas and developing regions~\cite{ITU_Hub_2023}. 
This gap contributes to a persistent digital divide, placing those without access to the Internet and other information and communication technologies (ICTs) at a marked socio-economic disadvantage.
Although various efforts have been made to improve rural connectivity through traditional backhaul infrastructure, these solutions often face significant challenges~\cite{yaacoub2020key}. In many cases, the geographic and economic conditions of rural areas make it difficult to deploy and maintain such infrastructure, further widening the access gap~\cite{yaacoub2020key}.
\par
Delay Tolerant Networking (DTN) has emerged as a viable solution for extending connectivity to rural and underserved areas~\cite{perumal2022enhanced}. Operating on a store-carry-forward paradigm, DTN allows data to be stored locally on mobile devices, transported as devices move, and forwarded when they enter the communication range of another device. 
This type of DTNs, known as mule-based DTNs, enables direct device-to-device communication using local radio signals, which reduces the need for fixed network infrastructure.
DTN is particularly suited for resource-constrained environments, offering a cost-effective backhaul-like connectivity approach to bridging the digital divide where traditional connectivity infrastructure may be impractical or costly~\cite{DTN_Abdeljabar2025}. This communication paradigm makes use of existing transportation systems, such as buses and trucks, to physically carry data between network points. These mobile carriers serve as the primary mechanism for data movement in the absence of continuous network coverage. 
\par
Informal public transportation systems offer a promising opportunity for supporting DTNs in developing regions. One such system is the use of minibus taxis, which play a central role in the daily transportation of people in many developing countries.
For example, in South Africa, medium-sized minibus taxis dominate the public transport sector, with over 60~\% of the population relying on them for mobility~\cite{ndibatya2014modelling}. These taxis transport approximately 14 million passengers daily, proving their significance within the transportation network. 
When used as mobile data mules, these vehicles can carry information between rural areas that lack Internet access and urban centres that are connected to the Internet. Such integration of mobility and data transfer offers a practical and cost-effective way to enhance connectivity in regions where traditional infrastructure is limited or absent.

%%%%%%%%%%%%%%%%%%%%%%%%%%%%%%%%%%%%%%%%%%%%%
%%%%%%%%%%%%%%%%%%%%%%%%%%%%%%%%%%%%%%%%%%%%%
\subsection{Related Works}
%%%%%%%%%%%%%%%%%%%%%%%%%%%%%%%%%%%%%%%%%%%%%
%%%%%%%%%%%%%%%%%%%%%%%%%%%%%%%%%%%%%%%%%%%%%
\par
DTN was originally developed by NASA for interplanetary satellite communication; however, it later showed significant potential in extending connectivity to rural regions where traditional communication infrastructure is unavailable or costly. DTN enables intermittent and opportunistic access to the Internet through a store, carry, and forward mechanism, allowing data to be transferred whenever communication opportunities arise.
Several studies have investigated the application of DTN in rural connectivity projects across different parts of the world. In~\cite{rahman2013delay}, authors proposed the use of public transport buses to carry data and establish intermittent links between remote villages in Bangladesh, effectively overcoming infrastructural barriers. Similarly, a study in South Africa~\cite{galati2014mobile} examined the use of taxis and other transport modes to facilitate data transfer between rural communities. Additional studies have emphasized DTNs flexibility in addressing geographic constraints. 
For example, the authors of~\cite{grasic2014revisiting} explored DTN deployments in remote mountainous areas of Sweden, where helicopters were used as mobile data mules. In another case, the authors of~\cite{ntareme2011delay} demonstrated DTNs adaptability through the use of Android-based devices to support communication in resource-constrained environments.
Taken together, these studies demonstrate that DTN can adapt to a wide range of rural environments and transportation contexts, offering a flexible and cost-effective solution for addressing connectivity challenges in underserved areas.
\par
While previous studies have demonstrated the feasibility of DTN systems for rural connectivity using mobile data mules such as buses, ferries, and helicopters~\cite{rahman2013delay, galati2014mobile, grasic2014revisiting, ntareme2011delay, pentland2004daknet, guo2007very, luo2018internet}, many of these works remain primarily empirical or descriptive, offering limited formal analysis of system performance. In particular, the integration of informal public transport systems, such as the ones found in many cities in developing countries, into DTN framework is still underexplored, especially with respect to quantifying key communication performance metrics. This paper addresses these gaps by introducing a probabilistic modeling framework based on renewal theory to capture the stochastic movement patterns of informal transport systems and incorporate them in the DTN design. Analytical expressions are derived for performance metrics such as the Mean Peak Age of Information (MPAoI), and cost considerations are incorporated to determine the minimum number of DTN data mules required to meet performance targets. The framework is validated using real-world mobility datasets from cities including Nouakchott, Accra, and Addis Ababa, reflecting the irregularities in informal transport behavior. This integrated approach of formal modeling, analytical evaluation, and validation with real-world mobility traces distinguishes this work from prior efforts and provides a more systematic basis for deploying mule-based DTNs in rural environments.

%%%%%%%%%%%%%%%%%%%%%%%%%%%%%%%%%%%%%%%%%%%%%
%%%%%%%%%%%%%%%%%%%%%%%%%%%%%%%%%%%%%%%%%%%%%
\subsection{Article Outline}
%%%%%%%%%%%%%%%%%%%%%%%%%%%%%%%%%%%%%%%%%%%%%
%%%%%%%%%%%%%%%%%%%%%%%%%%%%%%%%%%%%%%%%%%%%%
\par
Motivated by the pressing need to bridge the digital divide, this work investigates the performance of DTN frameworks that leverage informal public transportation systems as data mules. We propose a general methodology to evaluate DTN performance from a communication perspective, with particular focus on optimizing the number of DTN data mules required to meet communication performance targets.
The primary research question explored in this article concerns the effectiveness of such systems in delivering Internet-like connectivity to underserved or remote regions. Specifically, we study scenarios where two disconnected areas, an Internet-connected urban center and a rural location with limited or no connectivity, are linked through the movement of DTN data mules. Our aim is to assess the quality of this connection by analyzing data flow and evaluating performance metrics.
\par
To this end, we develop a probabilistic model based on renewal theory to describe the mobility behavior of informal transport systems, including transportation routes, travel durations and waiting times. 
We also incorporate cost and quality-of-service (QoS) constraints into the design process. Particularly, we consider the MPAoI and average data transmission rate as key QoS metrics, drawing parallels to optimization strategies used in traditional cellular networks, but adapted to the mobile nature of DTN nodes. 
The framework is validated through real-world mobility datasets from cities including Nouakchott, Accra, and Addis Ababa, capturing the irregular dynamics of informal transport. Our goal is to provide a scalable framework that provides indicative metrics for the predicted communication performance that such systems can achieve, and hence offering insights into their practical feasibility and design optimization in rural contexts.
The main contributions of this article are summarized as follows:
\begin{itemize}
    \item We introduce a comprehensive framework to model the mobility patterns of informal transport systems, including travel durations and waiting times at bus stops. This framework is based on renewal theory to characterize data mule movement between rural and urban regions.
    \item We analyze key performance metrics such as the mean transmitted data size, average data transmission rate, and Mean Peak Age of Information (MPAoI). Although exact analytical expressions for MPAoI is complex, we provide an analytical approximation of the MPAoI based on the superposition of renewal processes at equilibrium, and validate the approximation via simulations. 
    \item We validate the proposed framework using real-world mobility datasets from informal public transport systems in cities including Nouakchott, Accra, and Addis Ababa. These case studies demonstrate the practical applicability and scalability of our framework.
\end{itemize}

\subsection{Notation Summary}
To facilitate understanding and maintain consistency throughout this manuscript, Table~\ref{tab:notation} provides a comprehensive summary of all key notations, symbols, and variables used in our analysis. The notations are organized into several categories: system parameters that define the network configuration, random variables that characterize the stochastic behavior of the transport system, performance metrics that quantify system effectiveness, and derived quantities used in the analytical framework.
\begin{table}[!h]
\captionsetup{}
\caption{Summary of Notations}
\label{tab:notation}
\centering
\renewcommand{\arraystretch}{1.1}
\begin{tabular}{@{}p{0.6cm}p{7.5cm}@{}}
\toprule
\textbf{Notation} &  \quad \textbf{ Description} \\
\midrule
\multicolumn{2}{@{}l}{\textit{System Parameters}} \\
$N$ & Number of vehicles serving the route between city and village \\
$n$ & Number of vehicles equipped with DTN modules ($n \leq N$) \\
$R_{\text{link}}$ & Wireless link data rate between a DTN-equipped vehicle and a fixed DTN gateway during contact time (Mbps) \\
$\mathcal{R}$ & Effective DTN uplink/downlink data transmission rate (Mbps)  \\
$c_1$ & Minimum contact time at bus stops (minutes) \\
$c_2$ & Maximum contact time at bus stops (minutes) \\
$T_{\min}$ & Minimum travel time between regions A and B (minutes) \\
$C$ & Cost of a single DTN node placed on vehicles (USD) \\
$C_{\text{gateway}}$ & Cost of a fixed DTN gateway (USD) \\
\midrule
\multicolumn{2}{@{}l}{\textit{Random Variables}} \\
$T_{c,A}$ & Contact time at region A (urban gateway), $T_{c,A} \sim \text{Unif}(c_1, c_2)$ \\
$T_{c,B}$ & Contact time at region B (rural gateway), $T_{c,B} \sim \text{Unif}(c_1, c_2)$ \\
$T_d$ & Random delay during vehicle's travel, $T_d \sim \text{Exp}(1/\overline{T}_{d}))$ \\
$\overline{T}_{d}$ & Average travel time delay, $\mathbb{E}[T_d]$ \\
$T_{AB}$ & Travel time from region A to region B, $T_{AB} = T_{\min} + T_d$ \\
$T_{BA}$ & Travel time from region B to region A, $T_{BA} = T_{\min} + T_d$ \\
$T_{v,i}$ & Inter-arrival time of the $v$-th vehicle for its $i$-th round trip \\
$T_s$ & Inter-arrival time of the superimposed process (all $n$ vehicles) \\
$m_v$ & Data size transmitted or received per vehicle arrival (Mbits) \\
\midrule
\multicolumn{2}{@{}l}{\textit{Performance Metrics}} \\
AoI & Age of Information \\
MAoI & Mean Age of Information \\
PAoI & Peak Age of Information\\
MPAoI & Mean Peak Age of Information \\
$\mathbb{E}[\mathcal{R}]$ & Mean data transmission rate of the DTN system (Mbps) \\
$\mathbb{E}[m_v]$ & Mean transmitted data size per vehicle arrival (Mbits) \\
$\lambda_s$ & Arrival rate of the superimposed process (vehicles per minute) \\
\midrule
\multicolumn{2}{@{}l}{\textit{Derived Quantities}} \\
$N_v(t)$ & Renewal process for the $v$-th vehicle \\
$N_s(t)$ & Superimposed renewal process for all $n$ DTN-equipped vehicles \\
$\mathbb{E}[T_{v,i}]$ & Mean inter-arrival time of a single vehicle (mean round-trip time) \\
$\mathbb{E}[T_s]$ & Mean inter-arrival time of the superimposed process \\
$F_X(x)$ & Cumulative distribution function (CDF) \\
$f_X(x)$ & Probability density function (PDF) \\
$t_l$ & Generation time of the $l$-th update at the source \\
$t'_l$ & Reception time of the $l$-th update at the monitor \\
$u(t)$ & Timestamp of the most recent update received by time $t$ \\
$\mu$ & Mean round-trip time, $\mathbb{E}[T_{c,A} + T_{AB} + T_{c,B} + T_{BA}]$ \\
\bottomrule
\end{tabular}
\end{table}

%%%%%%%%%%%%%%%%%%%%%%%%%%%%%%%%%%%%%%%%%%%%%
%%%%%%%%%%%%%%%%%%%%%%%%%%%%%%%%%%%%%%%%%%%%%
%%%%%%%%%%%%%%%%%%%%%%%%%%%%%%%%%%%%%%%%%%%%%
\section{System Model}\label{section:system_model}
%%%%%%%%%%%%%%%%%%%%%%%%%%%%%%%%%%%%%%%%%%%%%
%%%%%%%%%%%%%%%%%%%%%%%%%%%%%%%%%%%%%%%%%%%%%
%%%%%%%%%%%%%%%%%%%%%%%%%%%%%%%%%%%%%%%%%%%%%

%%%%%%%%%%%%%%%%%%%%%%%%%%%%%%%%%%%
%%%%%%%%%%%%%%%%%%%%%%%%%%%%%%%%%%%
\subsection{Overview}
%%%%%%%%%%%%%%%%%%%%%%%%%%%%%%%%%%%
%%%%%%%%%%%%%%%%%%%%%%%%%%%%%%%%%%%
\par
In this section, we present a model for public transportation systems that connect two distant regions, an urban city and a rural village. Broadly, public transport systems can be classified into two categories. The first includes formal transport systems that operate on fixed schedules and predefined routes, often regulated by governmental or municipal authorities. These are commonly found in urban areas. The second category comprises informal systems, which are typically unregulated and lack predefined schedules or fixed routes. Examples include minibuses, shared taxis, and similar modes of transport frequently used in developing countries~\cite{mittal2024efficient}.
In informal transport systems, vehicles often wait at central stops until a sufficient number of passengers with similar destinations are gathered before departing. The departure times and the routes taken are not fixed in advance, and both the waiting time at stops and the travel time between regions are inherently random. This work focuses on such systems, modeling their stochastic behavior through a probabilistic framework. The framework captures the key sources of randomness and provides analytical tools to support the design and deployment of DTNs using these transportation systems as data mules.

%%%%%%%%%%%%%%%%%%%%%%%%%%%%%%%%%%%
%%%%%%%%%%%%%%%%%%%%%%%%%%%%%%%%%%%
\subsection{Considered Scenario}
%%%%%%%%%%%%%%%%%%%%%%%%%%%%%%%%%%%
%%%%%%%%%%%%%%%%%%%%%%%%%%%%%%%%%%%
\par
\begin{figure}[h!]
    \centering
    \includegraphics[width=1\linewidth]{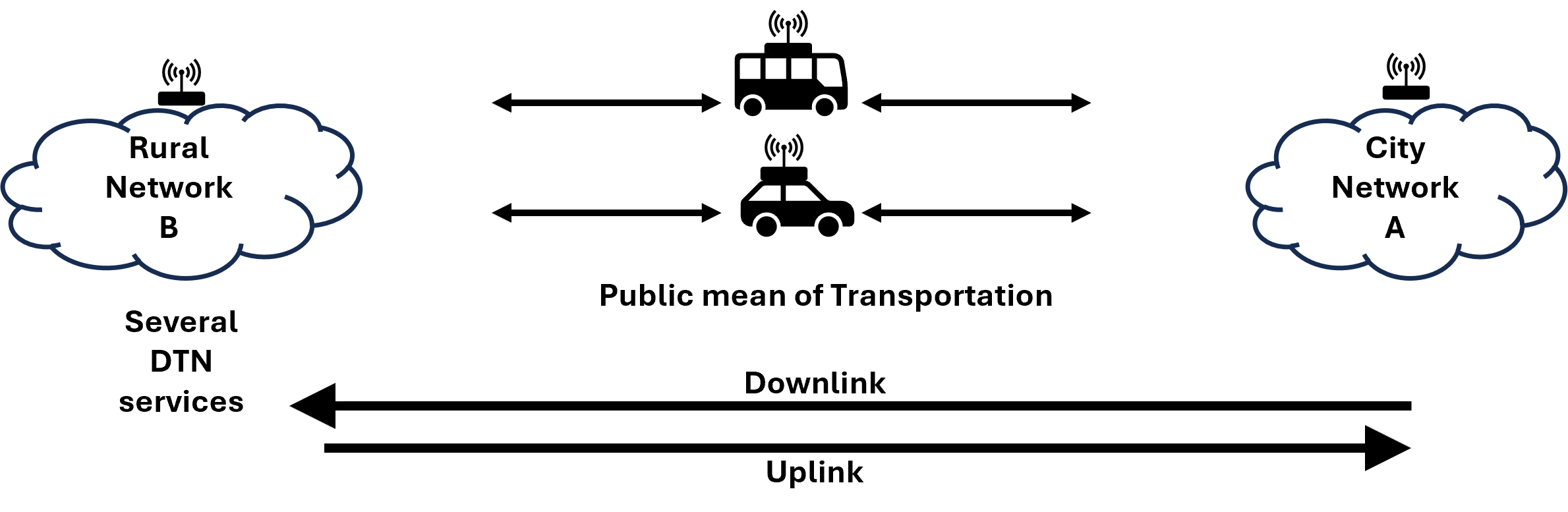}
    \caption{The considered DTN scenario, where DTN modules carried via informal public transport means are used to connect the city and rural networks.}
    \label{fig:DTN_Scenario}
\end{figure}

We consider a DTN system that operates between two geographically separated regions: an urban area with Internet connectivity and cloud services, and a rural village with limited or no Internet infrastructure. 
Fig.~\ref{fig:DTN_Scenario} illustrates the architecture we consider in this work.
In the rural region, various DTN-based services operate under the assumption of non-real-time communication, such as asynchronous email, web caching, digital learning platforms~\cite{DTN_Abdeljabar2025}, and IoT-based environmental or agricultural monitoring. These services rely on intermittent but reliable data delivery, making DTN a suitable solution in such contexts.
Connectivity between the two regions is facilitated by informal public transportation vehicles, such as minibuses or shared taxis, which travel between a designated bus stop in the city and another in the rural area. Among the total fleet of $\mathcal{N}$ vehicles serving this route, only $n$ ($n \leq \mathcal{N}$) are equipped with DTN communication modules. Each region's main bus stop is equipped with a DTN gateway that establishes wireless links (e.g., via Wi-Fi) with arriving vehicles. 

For simplicity, these gateways, referred to as node A in the urban region (region A) and node B in the rural region (region B), serve as fixed communication endpoints for data transfer. As DTN-equipped vehicles move between regions A and B, they carry data in both directions: from the Internet (via node A) to rural users (via node B), and from rural users back to Internet-based services. These vehicles thus function as mobile data mules, bridging the connectivity gap between urban and rural regions.
The DTN architecture described in this work is particularly relevant for regions where informal public transportation systems dominate intercity mobility, which are prevalent in many cities in developing countries~\cite{mittal2024efficient}.

\par
For simplicity and analytical tractability, the proposed model considers communication only between DTN-equipped vehicles and the fixed DTN gateways (nodes A and B), excluding vehicle-to-vehicle (V2V) communication. In the context of informal rural-urban transport, such vehicles typically travel long distances on disjoint schedules and with random corridors, making in-transit V2V encounters infrequent. Moreover, informal systems often operate sequentially, with vehicles departing only after a sufficient number of passengers are on board, further reducing the likelihood of overlap between DTN-equipped vehicles during transit.
It should be noted that the discussed DTN framework targets applications that do not require instant data delivery. Some IoT scenarios fit this category, including environmental monitoring, sensor data collection in agriculture, and wildlife tracking, where measurements are stored and forwarded when possible~\cite{benhamida2019toward, li2020resource, bounsiar2019enable}. However, the framework is not intended for IoT services that need continuous or low-latency connections, such as critical emergency alerts, industrial automation, or real-time health monitoring. Such services must rely on alternative connectivity solutions that guarantee immediate and continuous data transmission.

%%%%%%%%%%%%%%%%%%%%%%%%%%%%%%%%%%%%%%%%%%%%%%%%%%%%%%%%%%%%
%%%%%%%%%%%%%%%%%%%%%%%%%%%%%%%%%%%%%%%%%%%%%%%%%%%%%%%%%%%%
\subsection{DTN Data Mule Model}\label{sec:Data_mule_model}
%%%%%%%%%%%%%%%%%%%%%%%%%%%%%%%%%%%%%%%%%%%%%%%%%%%%%%%%%%%%
%%%%%%%%%%%%%%%%%%%%%%%%%%%%%%%%%%%%%%%%%%%%%%%%%%%%%%%%%%%%
\par
In this part, we develop a probabilistic model that describes the DTN data mules movement traversing between the two regions. We assume that these vehicles are mainly informal public transport providing shared transportation for multiple passengers, such as medium-sized minibus taxis and vans that dominate the public transport sector in many developing countries~\cite{ndibatya2014modelling}. These vehicles operate without fixed, predefined intermediate stops or schedules but with roughly fixed service corridors that start/end at a main bus stop at each region~\cite{mittal2024efficient}. 
Therefore, the time it takes for each vehicle to travel between the two regions and the waiting time at the main bus stop are both random. 
\textcolor{black}{In addition, a key metric of interest in our framework is the time interval between successive DTN-equipped vehicles' arrivals at each bus stop, as it directly impacts the data transmission rate and the effectiveness of the DTN system. The following subsections provide detailed modeling of these timing distributions and explain how they influence communication performance. }

%%%%%%%%%%%%%%%%%%%%%%%%%%%%%%%%%%%%%%%%%%%%%%%%%%%%%%%%%%%%%%%%%%
\subsubsection{\textbf{Contact Time}} \label{section:Contact_time}
%%%%%%%%%%%%%%%%%%%%%%%%%%%%%%%%%%%%%%%%%%%%%%%%%%%%%%%%%%%%%%%%%%
A vehicle arrives at the main bus stop and waits for passengers to board or alight from the vehicle. The waiting time for the vehicles varies depending on the time of the day and the availability of customers to travel to the intended destination. 
During this time, a DTN-equipped vehicle establishes a wireless connection with the DTN gateway located at the bus stop and starts to send/receive data packages with data rate \(R_{\text{link}} \) (in Mbps, representing the wireless link data rate between DTN-equipped vehicles and DTN gateways). 
We call this period the \textit{contact time} between the vehicles and DTN gateway, and it determines the data size sent/received for each time the vehicles stop at the bus stop. In practice, the waiting time is random in nature and could be few minutes or tens of minutes; hence, we model the contact time, denoted as $T_c$, for each vehicle stopping near DTN gateways at each region (nodes A and B, see Fig \ref{fig:DTN_Scenario}) as uniform distribution $T_{c, \{A, B\}} \sim Unif(c_1,c_2)$, where $c_1$ and $c_2$ are the minimum and maximum observed waiting time for drivers at the bus stops, respectively. 
This typically captures the randomness of the time needed for passengers to board and alight from the vehicle and/or the drivers to take a break after each trip. 
In addition, the uniform distribution also assumes no inherent bias within the range, which reflects the variability caused by operational factors, and provides a straightforward and effective representation of the randomness observed in practice. \textcolor{black}{We note that more sophisticated distributions could potentially offer greater accuracy in specific contexts, particularly in highly dense or disrupted environments. However, such refinements would require extensive empirical data collection on actual waiting time patterns. Importantly, our analytical framework focuses on the expected value of system delays, where the mean contact time plays a dominant role. This focus ensures that our key conclusions remain robust regardless of whether the underlying distribution is uniform or follows alternative forms with comparable mean values.}

%%%%%%%%%%%%%%%%%%%%%%%%%%%%%%%%%%%%%
\subsubsection{\textbf{Travel Time}}\label{section:Travel_Time}
%%%%%%%%%%%%%%%%%%%%%%%%%%%%%%%%%%%%%
Once passengers board or alight from the vehicle, it starts to travel to the other region. In practice, during the journey, the drivers often adjust their route and stop at intermediate locations dynamically, for example, to find more customers, circumvent traffic, or avoid police checkpoints~\cite{mittal2024efficient}. Therefore, we model the time to travel from region A to B (and from B to A), denoted as $T_{\{AB, BA\}}$, to be a combination of two parts; $T_{min}$, which is the minimum time needed for a vehicle to travel between the city and village (following an optimum path, and without considering any delay sources), and $T_{d}$, a delay that captures the drivers' intermediate stopping, traffic encountered during its journey, and any other sources of random delay. $T_{min}$ is constant and calculated based on the distance and optimum route between the city and the village, while $T_{d}$ is modeled as exponentially distributed random variable $T_{d} \sim Exp(1/\overline{T}_{d})$, where $\overline{T}_{d}$ is the average delay of the travel time of the vehicle.
We model the delay \(T_d\) as an exponentially distributed random variable because the exponential distribution is commonly used to capture random delays, reflecting the dynamic and unpredictable nature of informal transport systems, where delays arise due to independent factors such as intermediate stops, traffic, or detours. 
The exponential model also provides mathematical simplicity, allowing analytical traceability while investigating communication metrics for DTNs.  
Hence, $T_{\{AB, BA\}} = T_{min} + T_{d}$, which means that  $T_{\{AB, BA\}}$ is a shifted-exponentially distributed random variable with mean delay $T_{min} + \overline{T}_{d}$.

%%%%%%%%%%%%%%%%%%%%%%%%%%%%%%%%%%%%%%%%%%%%%%%%%%%%%%%%%%%%%%%%%%%%%%%%%%%%%%%%%%%%%%%
\subsubsection{\textbf{Inter-arrival time of vehicles}}\label{sec:inter-arrival_times}
%%%%%%%%%%%%%%%%%%%%%%%%%%%%%%%%%%%%%%%%%%%%%%%%%%%%%%%%%%%%%%%%%%%%%%%%%%%%%%%%%%%%%%%
\par
In our scenario, we consider $\mathcal{N}$ vehicles serving passengers between the city and village, of which $n$ vehicles ($n \leq \mathcal{N}$) are equipped with DTN modules to deliver data packages. These vehicles traverse between the two regions throughout the day, with round-trip times modeled as described in Sections \ref{section:Contact_time} and \ref{section:Travel_Time}.
One key metric in this analysis is the inter-arrival time of vehicles at each region, which directly impacts data transmission rates and communication metrics. 
Here, we define the \emph{inter-arrival} time of a single vehicle as the time between two successive arrivals of the same DTN-equipped vehicle at a region (e.g., region A). This is equivalent to the vehicle's full round-trip time, encompassing its contact time at region A, travel to region B, contact time at region B, and return to region A.
To model this, we represent each vehicle movement (the $v$-th vehicle) as a renewal process, denoted as $N_v(t)$, with independent and identically distributed inter-arrival times. The $i$-th inter-arrival time of the $v$-th vehicle is donated by $T_{v,i}$. For a vehicle starting at any region (region A or B), the time to complete successive round trips follows the renewal pattern $T_{v,1}, T_{v,2}, \ldots$, with each renewal time accounting for both travel and waiting durations. 
In our considered scenario, the inter-arrival time $T_{v,i}$ is given as the time needed for the vehicle to complete a round-trip starting from one of the regions, 
Formally, we define the inter-arrival time of a DTN-equipped vehicle $v$ as the total time taken to complete one full round-trip starting at node A (urban), travelling to node B (rural), and returning to node A. This includes the contact time at node A ($T_{c,A}$), the travel time from A to B ($T_{AB}$), the contact time at node B ($T_{c,B}$), and the return travel time from B to A ($T_{BA}$). Hence, the $i$-th inter-arrival time for the $v$-th DTN-equipped vehicle is expressed as:
\begin{equation}\label{equ:inter-arrival_time}
    T_{v,i} = T_{c,A} + T_{AB} + T_{c,B} + T_{BA}
\end{equation}
where each term in the right-hand side of Equation~\eqref{equ:inter-arrival_time} corresponds to the $v$-th vehicle during its $i$-th round trip, and the subscripts \(v\) and \(i\) are omitted for notational simplicity because the random variables \((T_{c,A}, T_{AB}, T_{c,B}, T_{BA})\) are assumed independent and identically distributed (i.i.d.) across vehicles and trips, with the understanding that each is implicitly indexed by \((v,i)\) in analysis and simulation.
To evaluate the aggregate performance of the system with $n$ DTN-equipped vehicles, we extend the above analysis to include all $n$ vehicles.
Suppose that the travel time of each vehicle traversing between the two regions is independent of the others. We model the arrival time of $n$ vehicles at region A (or B) as a superposition of $n$ independent renewal processes, denoted by $N_s(t)$ and given as follows:
\begin{equation} \label{Equ:superimposed_process}
    N_s(t) = N_1(t) + N_2(t) + \dots + N_n(t)
\end{equation}
$N_s(t)$ in Equation~\eqref{Equ:superimposed_process} describes the superimposed arrival process from $n$ vehicles as seen at each  region, i.e., the $i$-th inter-arrival time of the superimposed process $N_s(t)$, denoted as $T_{s,i}$, represents the time elapsed between the arrival of any consecutive vehicles at both regions.
In general, the superposition of independent renewal processes is not a renewal process, and finding the renewal time (inter-arrival time of aggregated vehicles) distribution of the superimposed process is not straightforward \cite{cox1962renewal}. Nevertheless, we focus on limiting distribution for intervals remote from the time origin, or equivalently, we look at the component process as an equilibrium renewal process. In other words, we assume that the vehicles have been running long before it is first observed. The assumption of observing the process at equilibrium is reasonable because vehicles are typically in continuous operation over the day, and the DTN is likely to have reached a steady state by the time it is first observed. 
Evidently, assuming the superimposed process is observed at equilibrium allows for direct computation of the mean inter-arrival time of the superimposed process. 
To clarify, suppose that the mean inter-arrival time of each of the individual processes is given by $\mathbb{E}[T_{v,i}] = \mathbb{E}[T_{v}]= \mu$ (the $i$ subscript was omitted as all inter-arrival times are assumed to be identically distributed across vehicles and round trips). It was shown in \cite{cox1962renewal} that, at equilibrium, the mean inter-arrival time of the superimposed process in Equation~\eqref{Equ:superimposed_process}, is given by:
\begin{equation} \label{Equ:Mean_InterArrivalTime}
    \mathbb{E}[T_{s, i}] = \mathbb{E}[T_{s}] = \frac{\mathbb{E}[T_{v}]}{n} = \frac{\mu}{n},
\end{equation}
with arrival rate, $\lambda_s$, of the superimposed process is: 
\begin{equation}\label{Equ:Mean_InterArrivalRate}
    \lambda_s = \frac{1}{\mathbb{E}[T_{s}]}=\frac{n}{\mu}
\end{equation}

This analytical relationship shown in Equation~\eqref{Equ:Mean_InterArrivalRate} not only confirms known intuitions but also enables a quantitative design framework for determining the minimum number of DTN-equipped vehicles, as to be discussed in the following section, required to meet specific communication performance goals, such as average data transmission rates and age of information thresholds, which is particularly useful to evaluate the effectiveness of DTNs.
In addition, the distribution of the inter-arrival time of the superimposed process in Equation~\eqref{Equ:superimposed_process}, denoted by $g(x)$, can be expressed as \cite{cox1962renewal}:
\begin{equation}
    g(x) = -\frac{d}{dx} \left[ \mathcal{F}(x) \left( \int_x^{\infty} \frac{\mathcal{F}(u)}{\mu} du \right)^{n-1} \right],
\end{equation}
where $\mathcal{F}(x)$ is the survival function of the inter-arrival time of the individual renewal process. 
In the above model, we assume that the DTN-equipped vehicles operate independently, and their inter-arrival times are drawn from independent and identically distributed renewal processes. This assumption simplifies the analysis and allows the derivation of closed-form expressions for key performance metrics. However, this assumption may not be applicable if the vehicles pass by congested urban environments or shared road segments, as vehicle delays may exhibit correlation due to common influences such as traffic jams, accidents, weather conditions, or scheduled events. Such dependencies could lead to bursty arrivals or prolonged service gaps, impacting the effectiveness of the DTN system. 
It is worth noting that modeling DTN-equipped vehicles as travelling independently is a reasonable approximation, particularly when the waiting times at bus stops are sufficiently long. In such cases, the variability introduced by independent boarding and alighting processes, along with driver behaviour, can help decouple successive trips, which further justifies our independence assumption.

%%%%%%%%%%%%%%%%%%%%%%%%%%%%%%%%
%%%%%%%%%%%%%%%%%%%%%%%%%%%%%%%%
%%%%%%%%%%%%%%%%%%%%%%%%%%%%%%%%
\section{Performance Metrics}
%%%%%%%%%%%%%%%%%%%%%%%%%%%%%%%%
%%%%%%%%%%%%%%%%%%%%%%%%%%%%%%%%
%%%%%%%%%%%%%%%%%%%%%%%%%%%%%%%%
In this section, we provide an analysis of the performance metrics of the considered mule-based DTN. We focus on two key performance metrics critical to DTNs: (i) the downlink/uplink data transmission rates and (ii) the Peak Age of Information (PAoI). These metrics provide insight into the efficiency and timeliness of data delivery, which are crucial for assessing the feasibility and scalability of mule-based DTNs.
We analyze each of these metrics to determine the minimum number of informal public transport vehicles that need to be equipped with DTN capabilities to meet specific performance requirements.
We begin by discussing the wireless link data rate between DTN-equipped vehicles and DTN gateways during the contact time. Subsequently, we analyze the transmitted data package size per-each contact time and the data transmission rate at each node. Finally, 
we analyze the PAoI and derive an analytical approximation for the Mean Peak Age of Information (MPAoI), which represents the expected value of PAoI across multiple update cycles.

%%%%%%%%%%%%%%%%%%%%%%%%%%%%%%%%%%%%%%%%%%%%%%%%%%%%%%%%%%%%
\subsection{Wireless Link Data Rate}\label{section:data_rate}
%%%%%%%%%%%%%%%%%%%%%%%%%%%%%%%%%%%%%%%%%%%%%%%%%%%%%%%%%%%%
\noindent This subsection examines the physical layer wireless link data rate between DTN-equipped vehicles and fixed DTN gateways during their contact time. As vehicles approach the bus stop, their radio terminals begin communicating with the fixed DTN gateway. Although affected by mobility factors with expected variable link quality, modern wireless technologies offer coverage ranging from a few meters to several hundred meters \cite{rani2022experimental, el2019wifi, rjab2024modeling}. The data rate improves as the vehicle and gateway move closer, significantly impacting the amount of data exchanged during the contact time.
While characterizing the transient period during a vehicle's approach to the gateway is challenging due to rapidly changing link quality, such as the signal-to-noise ratios (SNR), most data transmission occurs when the vehicle stops near the DTN gateway. 
For simplicity, we assume a fixed wireless link data rate \( R_{\text{link}}\) (in Mbps) during the vehicle's contact time, representing the peak physical-layer transmission capacity when the vehicle is stationary.
Additionally, we assume symmetrical uplink and downlink data rates during the contact time. This is a reasonable assumption, as modern Wi-Fi technologies share bandwidth between uplink and downlink transmissions using advanced multiple-access schemes \cite{baueroverview}.

%%%%%%%%%%%%%%%%%%%%%%%%%%%%%%%%%%%%%%%%%%%%%%%%%%%%%%%%%%
\subsection{Downlink/Uplink Data Transmission Rate}
%%%%%%%%%%%%%%%%%%%%%%%%%%%%%%%%%%%%%%%%%%%%%%%%%%%%%%%%%%
In DTN systems, communication often involves long delays and is designed for scenarios where continuous connectivity cannot be guaranteed. To account for these delays, we consider the total time a vehicle spends on its journey, including its contact time with fixed gateways, as part of the data transmission process. Consequently, the data transmission rate in such networks is defined differently, reflecting the intermittent nature of data exchanges between nodes. Specifically, data is transmitted in discrete chunks whenever DTN-equipped vehicle establishes contact with fixed DTN gateways.  
We define the effective data transmission rate, denoted as $\mathcal{R}$, as the total amount of data transmitted or received (in Mbits) for uplink or downlink over an observation period. Note that the transmission is not continuous, and only a portion of the observed period involves active data exchange. This metric provides a meaningful representation of network performance, enabling a direct analysis of the effect of vehicle density on data transmission rate. 
To quantify the downlink and uplink transmission rates for our scenario, we leverage the established inter-arrival time of vehicles described in Section \ref{sec:inter-arrival_times}. First, we calculate the average transmitted data size for each vehicle $v$, denoted as $m_v$, during the contact time for both uplink and downlink.  
\begin{lemma}[Mean Transmitted Data Size]\label{lemma:mean_transmitted_data_size}
The data size for uplink/downlink is proportional to the time needed to pick up the data package from region A (or B) and deliver it to region B (or A). Thus, the transmitted data size for the $v$-th vehicle during contact time is given by:  
\begin{equation}
    m_v = R_{\text{link}} \boldsymbol{\cdot} \min\{T_{c,A}, T_{c, B}\}
\end{equation}
where the complementary cumulative distribution function (CCDF) of the data size is:  
\begin{equation}
    F_m(m) = \left( \frac{c_2 - m/R_{\text{link}}}{c_2 - c_1} \right)^2, \quad m \in \left[c_1 R_{\text{link}}, c_2 R_{\text{link}}\right]
\end{equation}
where $c_1$ and $c_2$ are the minimum and maximum contact times at the bus stops, respectively, as defined in Section~\ref{section:Contact_time}. The mean transmitted data size for uplink/downlink is:  
\begin{equation}
    \mathbb{E}[m_v] = R_{\text{link}} \boldsymbol{\cdot} \left(\frac{2c_1 + c_2}{3}\right) \label{equ:mean_data_size}
\end{equation}
\\
Proof: See Appendix~\ref{section:Appendix_A}. \hfill $\blacksquare$
\end{lemma}
\noindent Given the known inter-arrival time of the aggregated vehicles, the mean effective data transmission rate can be expressed as:  
\begin{lemma}[Mean Data Transmission Rate]
For the mean data size per vehicle arrival, $\mathbb{E}[m_v]$, and vehicles arrival rate, $\lambda_s$, the mean data transmission rate, denoted as $\mathbb{E}[\mathcal{R}]$, is given by:  
\begin{equation}
    \mathbb{E}[\mathcal{R}] = \mathbb{E}[m_v]\lambda_s =\mathbb{E}[m_v] \boldsymbol{\cdot} \frac{n}{\mu}, 
\end{equation} 
\end{lemma}
This result shows that the mean data transmission rate is linearly proportional to the number of vehicles. This is consistent with Equation~\eqref{Equ:Mean_InterArrivalRate}, where the rate of vehicle arrivals is also linearly proportional to the number of vehicles.  

%%%%%%%%%%%%%%%%%%%%%%%%%%%%%%%%%%%%%%%%%%%%%%%%
\subsection{Peak Age of Information}
%%%%%%%%%%%%%%%%%%%%%%%%%%%%%%%%%%%%%%%%%%%%%%%%
\par
\textcolor{black}{The concept of Age of Information (AoI) was introduced to quantify the freshness of knowledge about a process observed from a distance \cite{kallitsis2022potential}. In other words, AoI measures the time elapsed since the last status update received from a source was generated. 
For clarity, assume a stochastic process $Y(t)$ is observed by a source node at the village, where samples of the process are extracted as time evolves. This process could represent the state of IoT devices, the status of a web-caching server, or any other service providing updates from the village. These samples, along with their timestamps $t_i$, are transmitted to a monitor node at the city network via DTN-equipped vehicles. The timestamp allows the monitor to track the freshness of the received samples. For example, if a sensor reading was created at time $t_g$ and arrives at the monitor at time $t_r$, the AoI at the moment of reception is $t_r - t_g$. After the update is received, the AoI grows linearly over time until a new update arrives and resets the age.}
\textcolor{black}{Unlike traditional delay metrics that measure how long it takes for a packet to travel from source to destination, AoI captures how timely the available information is, accounting for both delivery delay and the frequency of updates. This makes it especially relevant in DTN scenarios where updates may arrive sporadically due to intermittent connectivity. Lower AoI values indicate fresher information, which is important in applications with strict delay requirements.}
AoI can be mathematically defined as follows:

\begin{definition}[Age of Information] \label{Def:AoI}
Consider a communication system comprising a source that generates data updates and a monitor that receives these updates. Let $t_l$ denote the generation time of the $l$-th update at the source, and let $t'_l$ denote the reception time of the $l$-th update at the monitor. 
At any time $t$, let $u(t)$ represent the generation timestamp of the most recent update received by the monitor. Specifically, if the monitor has received updates up to the $\vartheta$-th update by time $t$, then $u(t) = t_\vartheta$, where $\vartheta = \max\{l : t'_l \leq t\}$.
The Age of Information (AoI) at time $t$, denoted as $A(t)$, is defined as the time elapsed since the generation of the most recently received update:
\begin{equation}
    A(t) = t - u(t).
\end{equation}
The AoI quantifies the freshness of information available at the monitor. When a new update is received at time $t'_l$, the AoI instantaneously drops to $A(t'_l) = t'_l - t_l$, which represents the end-to-end delay of that update. Between consecutive update arrivals, the AoI increases linearly with time, creating a sawtooth pattern. 
\end{definition}

\par
Another key metric of interest is the Peak Age of Information (PAoI), which describes the maximum AoI observed just before the arrival of a new update. PAoI can be considered as the upper bound of the AoI (worst-case data freshness) and is considered as the metric to measure the freshness of information in systems, as it typically provides more tractable solutions. Formally, PAoI is defined as:
\begin{definition}[Peak Age of Information] \label{Def:PAoI}
Following the definition of AoI in \ref{Def:AoI}, the PAoI when receiving the $l$-th update is given by:
\begin{equation}\label{Equ:PAoI}
    PAoI[l] = t_l^{'} - t_{l-1},
\end{equation}
where $t_l^{'}$ is the time the $l$-th update is received and $t_{l-1}$ is the generation time of the previous update.
\end{definition}
\par
\begin{figure}[h!]
    \centering
    \includegraphics[width=1\linewidth]{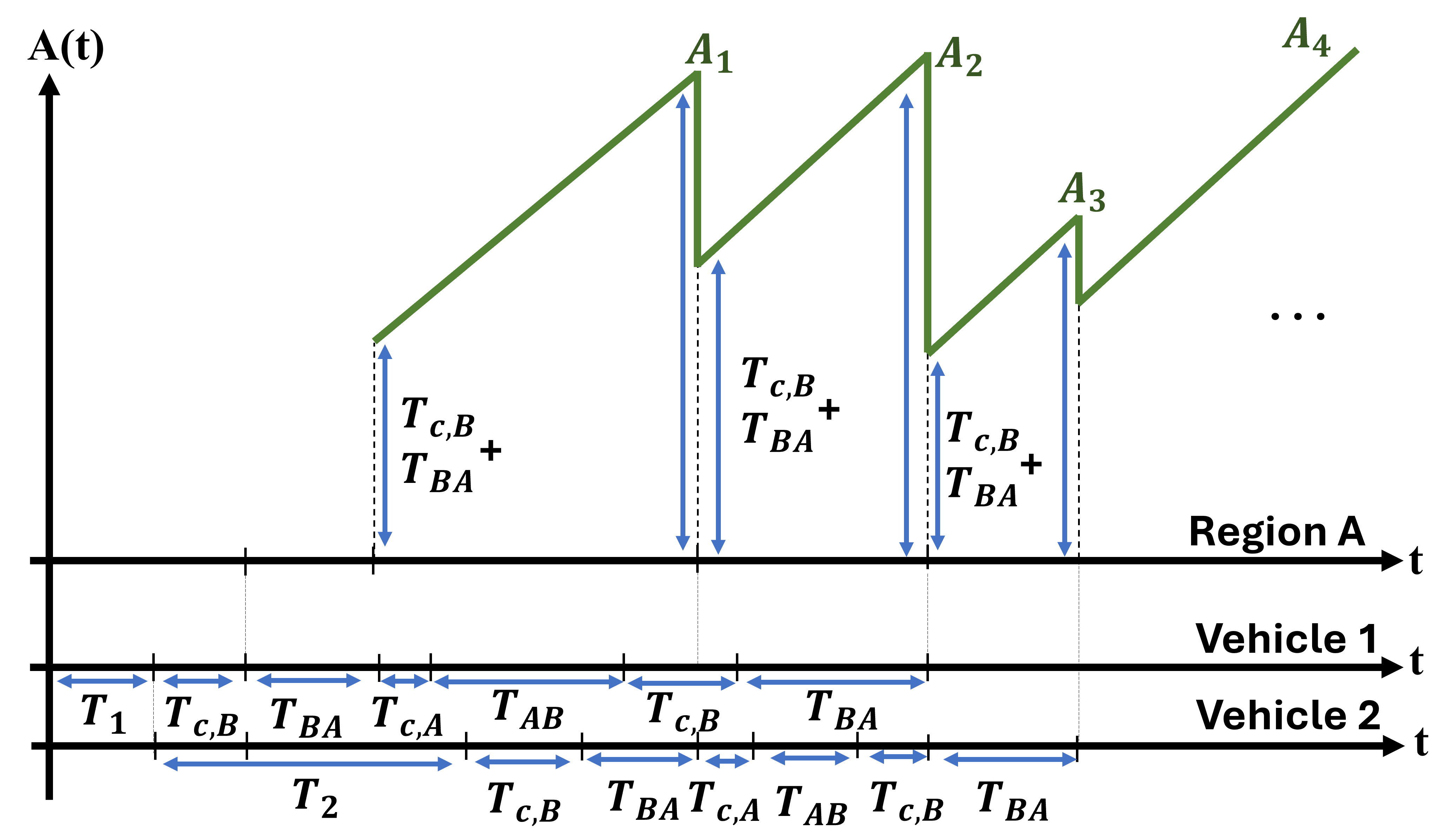}
    \caption{A sample realization of AoI and PAoI for a system with two DTN-equipped vehicles. The peaks $A_1, A_2, \text{\ldots }$  correspond to the PAoI values.}
    \label{fig:AoI_2-vehicles}
\end{figure}

Based on Definitions \ref{Def:AoI} and \ref{Def:PAoI}, a sample realization of the AoI in a system with two DTN-equipped vehicles operating between two regions illustrating the PAoI peaks ($A_1, A_2, ...$) from the perspective of a monitor node at the city receiving updates from the village is shown in Fig.~\ref{fig:AoI_2-vehicles}.
At any time $t$, the AoI grows linearly until a fresh update is delivered to the monitor node, where PAoI occurs. 

\par
To analyze the PAoI in our considered scenario, the following assumptions are made for the system under consideration. First, each DTN terminal (at the city, village, and mules) has sufficiently large buffers, ensuring no packet loss due to buffer overflow. This assumption is motivated by the availability of high-capacity and low-cost storage units. Second, the contact time during a vehicle’s stop at a node is long enough to allow the transmission of a complete update packet. This ensures that the updates are appropriately sized to fit within the contact time.
For a concrete example, we refer to Fig.~\ref{fig:AoI_2-vehicles} and consider observing the system starting at $t = T_1$, with vehicle 1 at region B and vehicle 2 arriving later at $t = T_1 + T_2$. Here, $T_1$ and $T_2$ are arbitrary initial times and do not affect the system's average performance. Vehicle 1 returns to region A after collecting data from region B at $t = T_1 + T_{c,B} + T_{BA}$. At this point, the AoI reaches $T_{c,B} + T_{BA}$. The AoI continues to grow until either vehicle 1 completes a round trip or vehicle 2 delivers a fresher update. In Fig.~\ref{fig:AoI_2-vehicles}, vehicle 2 arrives first, and the first PAoI is $A_1 = T_2 + T_{c,B} + T_{BA}$. The AoI then drops to $T_{c,B} + T_{BA}$ and resumes its linear growth until the next update arrives. 
\par
In general, obtaining a closed-form expression for the PAoI is challenging due to the inherent randomness of vehicle trip durations and the possibility that a vehicle may carry outdated updates if another vehicle delivers fresher data earlier. To make the analysis tractable, we assume that vehicles complete their trips sequentially, which aligns with typical operational patterns in informal public transport and is supported by the relatively long waiting times observed at main bus stops. Under this assumption, we leverage the results derived for the inter-arrival time in a system with $n$ DTN-equipped vehicles and observe that the PAoI depends primarily on three factors: the contact time at the rural bus stop ($T_{c,B}$), the travel time required for the data package to move between the rural village and the city ($T_{BA}$), and the difference in arrival times between two consecutive DTN-equipped vehicles, which follows the inter-arrival time distribution of the superimposed renewal process ($T_{s}$). Accordingly, the PAoI can be approximated as $PAoI \approx T_{c,B} + T_{BA} + T_{s}$. Additionally, the Mean PAoI (MPAoI), denoted as $\mathbb{E}[PAoI]$, is expressed as $\mathbb{E}[T_{c,B} + T_{BA} + T_s]$.  
We verify how close this approximation is to the actual quantity in the simulation section.
Moreover, as the number of vehicles increases, the MPAoI is expected to decrease and approach $T_{c,B} + T_{BA}$. Specifically, when the mean inter-arrival time of vehicles is reduced, the MPAoI follows. Thus, the MPAoI can be approximated as:
\begin{equation}
    MPAoI \approx  \mathbb{E}[T_{c,B} + T_{BA}] + \frac{\mu}{n}, \label{Equ:MPAoI_approximation}
\end{equation}
This relationship highlights the effect of deploying more DTN mule nodes at vehicles to minimize the system’s AoI (and PAoI).

%%%%%%%%%%%%%%%%%%%%%%%%%%%%%%%%%%%%%%%%%%%%
%%%%%%%%%%%%%%%%%%%%%%%%%%%%%%%%%%%%%%%%%%%%
\subsection{Data Traffic Requirements}
%%%%%%%%%%%%%%%%%%%%%%%%%%%%%%%%%%%%%%%%%%%%
%%%%%%%%%%%%%%%%%%%%%%%%%%%%%%%%%%%%%%%%%%%%
\par
In mule-based DTN systems, the amount of data exchanged between nodes primarily depends on the contact time during vehicle stops. Estimating the data traffic generated by users at the village is critical to optimizing system components, including storage capacities and wireless access points. DTN systems prioritize reliability over latency, favouring non-real-time applications such as email~\cite{lindgren2007experiences}, web-caching~\cite{naslund2013developing}, asynchronous messaging~\cite{grasic2011not}, and asynchronous digital learning platforms~\cite{DTN_Abdeljabar2025}. Additionally, DTNs can support IoT applications and collect sensor data from remote areas~\cite{benhamida2019toward, li2020resource, bounsiar2019enable, sarros2021intermittently}, where cost-effectiveness and scalability play a vital role for resource-constrained IoT ecosystems. As discussed, DTN systems are most effective when the data traffic does not depend on an immediate response. Suitable IoT examples include batch collection of sensor readings for agriculture, wildlife tracking, and long-term environmental monitoring. For these use cases, periodic uploads and opportunistic data transfer are sufficient. On the other hand, applications that require real-time communication or immediate action, such as critical alerts or industrial controls, fall outside the range of services supported by the mule-based DTN designs.
\par
The total data traffic generated at the village depends on the user density, denoted as $u$, and the traffic demand per user, denoted as $d$ in Mbits/day. Typical estimates suggest that email and asynchronous chat applications generate 10--50~Mbits/day per user, while digital learning platforms may produce hundreds of Mbits/day~\cite{penteldata_calculator}. Multimedia content requests (e.g., video streaming) generate significantly higher traffic, between 300--700~Mbits/hour, with general web browsing consuming around 50~Mbits/hour~\cite{omni_data_usage_calculator, att_data_calculator}. Based on these estimates, the daily data demand per user could reach approximately 1~Gbits/day, though this depends on specific DTN applications and user activity levels.
To meet the data traffic demands of a village, the total daily traffic, denoted as $\mathcal{D}_{\text{day}}$, can be calculated as:
\begin{equation}
    \mathcal{D}_{\text{day}} = u \cdot d \quad \text{(Mbits/day)}.
\end{equation}
From this, the required number of vehicles to be equipped with DTN can be estimated, considering their capacity and frequency of travel. This calculation informs decisions regarding the deployment and configuration of DTN infrastructure, ensuring sufficient resources for data exchange.

%%%%%%%%%%%%%%%%%%%%%%%%%%%%%%%%%
%%%%%%%%%%%%%%%%%%%%%%%%%%%%%%%%%
\subsection{Cost Considerations}
%%%%%%%%%%%%%%%%%%%%%%%%%%%%%%%%%
%%%%%%%%%%%%%%%%%%%%%%%%%%%%%%%%%
% \par
The proposed DTN architecture consists of two types of nodes: fixed DTN gateways located at urban and rural village bus stops and mobile DTN nodes (data mules) mounted on public transport vehicles. A typical DTN node comprises a processing unit (e.g., a small server or computer), a wireless adapter (e.g., Wi-Fi), a storage unit, a power supply with battery backup, and an enclosure for all components. The cost of each node ranges from a few hundred to over a thousand dollars depending on the hardware configuration \cite{grasic2014revisiting}. 
In practice, the cost of a DTN node depends heavily on its hardware configuration. Low-cost designs based on devices such as Raspberry Pi boards equipped with Wi-Fi dongles and microSD or SSD storage can bring the total cost to under a few hundred USD.
For mobile DTN nodes mounted on vehicles, power can be drawn directly from the vehicle's electrical system, eliminating the need for standalone battery systems or solar panels and thereby reducing both initial costs and maintenance requirements. This vehicle-powered approach is particularly practical since the nodes only need to operate during vehicle contact times.
On the other hand, 
fixed DTN gateways at rural locations without reliable grid power may require solar panels with battery backup to ensure continuous operation. Additionally, integrated systems with rugged enclosures, onboard batteries or solar panels, and pre-installed software stacks may raise the cost of the DTN terminals. 
Our framework is flexible and supports a wide range of cost parameters to accommodate this variability.
While connecting the urban DTN gateway to the Internet incurs additional expenses, most costs are one-time investments amortized over the network's operational lifespan.
In this DTN system, the primary variable cost is associated with deploying the mobile DTN nodes. To minimize these costs, we aim to determine the minimum number of mobile DTN nodes required to meet specific communication performance metrics. Assuming the cost of a single DTN node placed on vehicles is denoted as \( C \), and for the fixed DTN gateways denoted as \( C_{\text{gateway}} \), the total deployment cost can be formulated as an optimization problem while meeting QoS requirements, including MPAoI and mean data transmission rate. The problem is formulated as follows:
\begin{align}
    \min_{n} \quad & n \cdot C + 2 \cdot C_{\text{gateway}}, \label{equ:objective} \\
    \text{s.t.} \quad & \mathbb{E}[T_{c,B} + T_{BA}] + \frac{\mu}{n} \leq \mathbb{E}[\text{PAoI}]_{thr}, \label{equ:MPAoI_constraint} \\
                      & \mathbb{E}[m_v] \cdot \frac{n}{\mu} \geq \mathbb{E}[\mathcal{R}]_{thr}, \label{equ:Rate_constraint}
\end{align}
where \( \mathbb{E}[\text{PAoI}]_{thr} \) is the threshold for the MPAoI, and \( \mathbb{E}[\mathcal{R}]_{thr} \) is the required average data transmission rate. The constraints in Equations~\eqref{equ:MPAoI_constraint} and \eqref{equ:Rate_constraint} are to satisfy the QoS requirements presented by the MPAoI and mean data transmission rate, respectively. 
The optimal solution for the optimization problem with objective function in Equation~\eqref{equ:objective} and constraints in Equations~\eqref{equ:MPAoI_constraint} and~\eqref{equ:Rate_constraint} is straightforward. Let \( \alpha = \frac{\mu}{\mathbb{E}[\text{PAoI}]_{thr} - \mathbb{E}[T_{c,B} + T_{BA}]} \) and \( \beta = \frac{\mathbb{E}[\mathcal{R}]_{thr} \cdot \mu}{\mathbb{E}[m_v]} \). The minimum number of mobile DTN nodes required to satisfy both constraints at the same time minimizes the total cost is:
\begin{equation}
    n_{opt} = \left\lceil \max\{\alpha, \beta\} \right\rceil.
\end{equation}
This formulation provides a cost-effective approach to deploying a DTN system while ensuring that communication performance metrics are met.

%%%%%%%%%%%%%%%%%%%%%%%%%%%%%%%%%%%%%%%%%%%%%%%%
%%%%%%%%%%%%%%%%%%%%%%%%%%%%%%%%%%%%%%%%%%%%%%%%
%%%%%%%%%%%%%%%%%%%%%%%%%%%%%%%%%%%%%%%%%%%%%%%%
\section{Numerical Results}
%%%%%%%%%%%%%%%%%%%%%%%%%%%%%%%%%%%%%%%%%%%%%%%%
%%%%%%%%%%%%%%%%%%%%%%%%%%%%%%%%%%%%%%%%%%%%%%%%
%%%%%%%%%%%%%%%%%%%%%%%%%%%%%%%%%%%%%%%%%%%%%%%%
\par
In this section, we validate the investigated framework through numerical simulations. We first present results under various system parameters and demonstrate how closely the proposed MPAoI approximation aligns with simulation outcomes. Additionally, we utilize datasets from informal public transport systems in Nouakchott, Accra, and Addis Ababa, to evaluate system constraints in realistic scenarios.

\begin{table}[!h]
\caption{Simulation Parameters}
\label{tab:sim_params}
\centering
{
\begin{tabular}{@{}ll@{}}
\toprule
\textbf{Parameter} & \textbf{Value} \\
\midrule
$c_1$ & 3 minutes \\
$c_2$ & 5 minutes \\
$T_{\min}$ & 100 minutes \\
$\overline{T}_{d}$ & 20 minutes \\
$R_{\text{link}}$ & 10 Mbps \\
\bottomrule
\end{tabular}
}
\end{table}

\par
\begin{figure}[h!]
    \centering
    \includegraphics[width=1\linewidth]{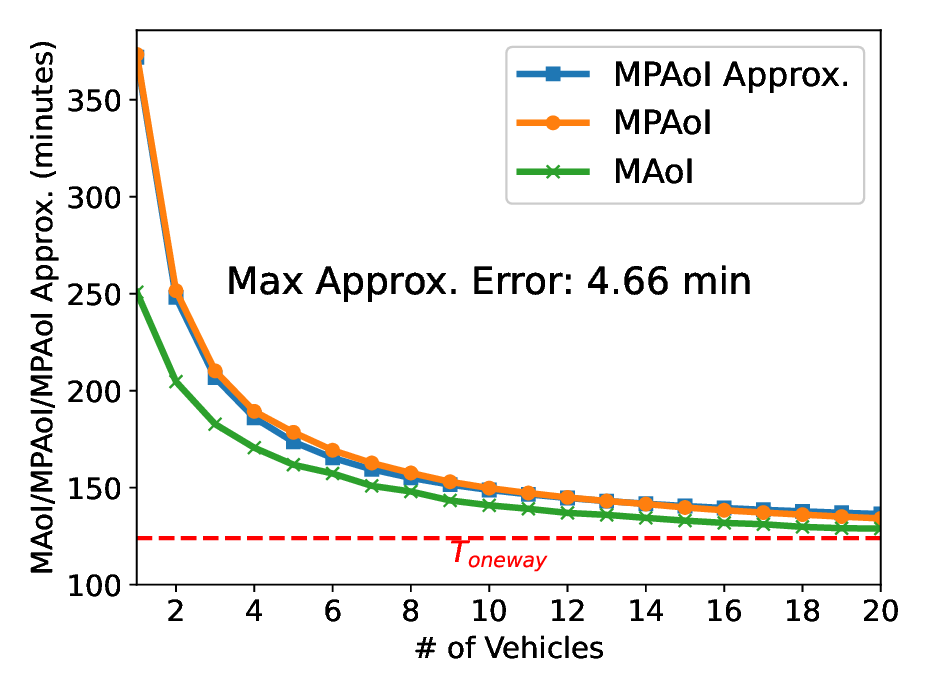}
    \caption{MAoI, MPAoI, and MPAoI approximation as described in Equation~\eqref{Equ:MPAoI_approximation}. The simulation was averaged over $10^5$ time units. The maximum approximation error represents the largest observed difference between the analytical approximation in Equation~\eqref{Equ:MPAoI_approximation} and the simulation-based MPAoI across all vehicle configurations tested (1 to 20 vehicles). The red dashed line represents the average time required for a vehicle to complete a one-way trip from the village to the city ($T_{oneway}$).}
    \label{fig:MPAoI_MAoI}
\end{figure}

To validate the MPAoI approximation derived in Equation~\eqref{Equ:MPAoI_approximation}, we developed a simulation framework based on the probabilistic model described in Section~\ref{sec:Data_mule_model}. The main simulation parameters are summarized in Table~\ref{tab:sim_params}, where we used the same waiting time parameters ($c_1$ and $c_2$) for regions A and B, and the same minimum delay parameter ($T_{\min}$) for both $T_{AB}$ and $T_{BA}$. The framework simulates discrete-time events over a maximum simulation duration of \( 10^{5} \) time units, with each unit representing one minute.
This is to ensure steady-state metrics of the considered scenario. The mobility of DTN-equipped vehicles in our simulations is modeled using the probabilistic framework described in Section~\ref{sec:Data_mule_model}, where round-trip durations are composed of random contact times (modeled as uniform distributions) and travel times (modeled as shifted exponential distributions). For validation, we incorporate real-world trace data from \cite{mittal2024efficient} to extract travel time statistics and compare performance across different urban-rural scenarios.
We calculate and plot the Mean AoI (MAoI), the MPAoI, and the approximated MPAoI (labeled ‘MPAoI Approx.’ in the figure), based on Equation~\eqref{Equ:MPAoI_approximation}. 
The results are shown in Fig.~\ref{fig:MPAoI_MAoI}. For a network with DTN-equipped vehicles ranging from 1 to 20, the maximum approximation error for the proposed MPAoI calculation is observed to be 4.66 minutes, demonstrating high accuracy for practical applications. 
The maximum approximation error represents the largest observed difference between the analytical approximation in Equation~\eqref{Equ:MPAoI_approximation} and the simulation-based MPAoI across all vehicle configurations tested (1 to 20 vehicles).
Additionally, in a network with a single DTN-equipped data mule, the MPAoI is approximately 375 minutes. However, as the number of data mules increases to 20, the MPAoI decreases to around 150 minutes. This substantial reduction shows the effectiveness of deploying additional DTN-equipped vehicles in improving data freshness and overall network efficiency. 
It can be also noted from Fig.~\ref{fig:MPAoI_MAoI} that, as the number of data mules increases, the MPAoI approaches its theoretical minimum (red dashed line in Fig.~\ref{fig:MPAoI_MAoI}), which corresponds to the average time required for a vehicle to complete a one-way trip from the village to the city, $T_{oneway}$, i.e., the age of the data package just as it arrives at the monitor at the city while generated $T_{oneway}$ minutes before at the village.

\par
\begin{figure}[h!]
    \centering
    \includegraphics[width=1\linewidth]{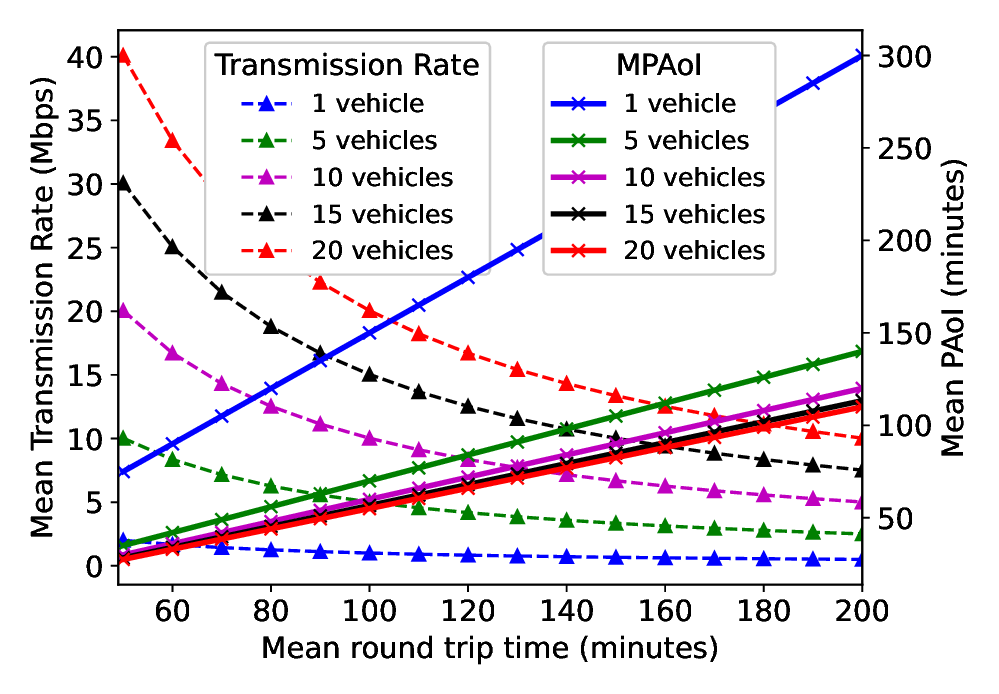}
    \caption{MPAoI and mean data transmission rate for different mean round-trip times and DTN-equipped vehicles in the network.}
    \label{fig:Round_tripVsMPAoI}
\end{figure}

Fig.~\ref{fig:Round_tripVsMPAoI} illustrates the impact of round-trip times on the mean data transmission rate and MPAoI. In this analysis, the mean minimum contact time at each bus stop is set to 10 minutes, and the wireless link data rate is assumed to be 10 Mbps. While modern Wi-Fi standards such as IEEE 802.11n/ac/ax can support data rates exceeding 600 Mbps~\cite{oughton2025future}, we conservatively assume a fixed average rate of 10 Mbps during the vehicle's contact time with DTN gateways. This reflects a more realistic data rate achievable in rural deployments where interference, energy limitations, and hardware constraints may affect link quality. The assumption accounts for average performance during the stationary phase of the vehicle's stop, when most data transmission occurs, and serves as a practical baseline for performance evaluation. Importantly, the analytical framework and simulation results remain valid for other data rate values, allowing adaptation to more bandwidth-constrained or unreliable rural environments.
Based on Equation~\eqref{equ:mean_data_size}, the mean data package size per vehicle arrival is around 6 GB.
From the figure, we observe that the transmission rate decreases exponentially as the mean round-trip time increases, consistent with the earlier analysis. Simultaneously, the MPAoI grows linearly with the round-trip time. Furthermore, as the number of DTN-equipped data mules increases, both the transmission rate and the MPAoI show significant improvement. For instance, in a network with a single DTN-equipped data mule and a mean round-trip time of 120 minutes, the transmission rate is approximately 2 Mbps, while the MPAoI is approximately 200 minutes. However, with 20 DTN-equipped data mules in the network, the transmission rate increases to around 20 Mbps, and the MPAoI drops to approximately 75 minutes. 
These results highlight that adding more data mules to the network improves the transmission rate linearly while reducing the MPAoI.
\par
\begin{figure}[h!]
	\centering
	\subfigure[]
	{\includegraphics[width=1.0\linewidth]{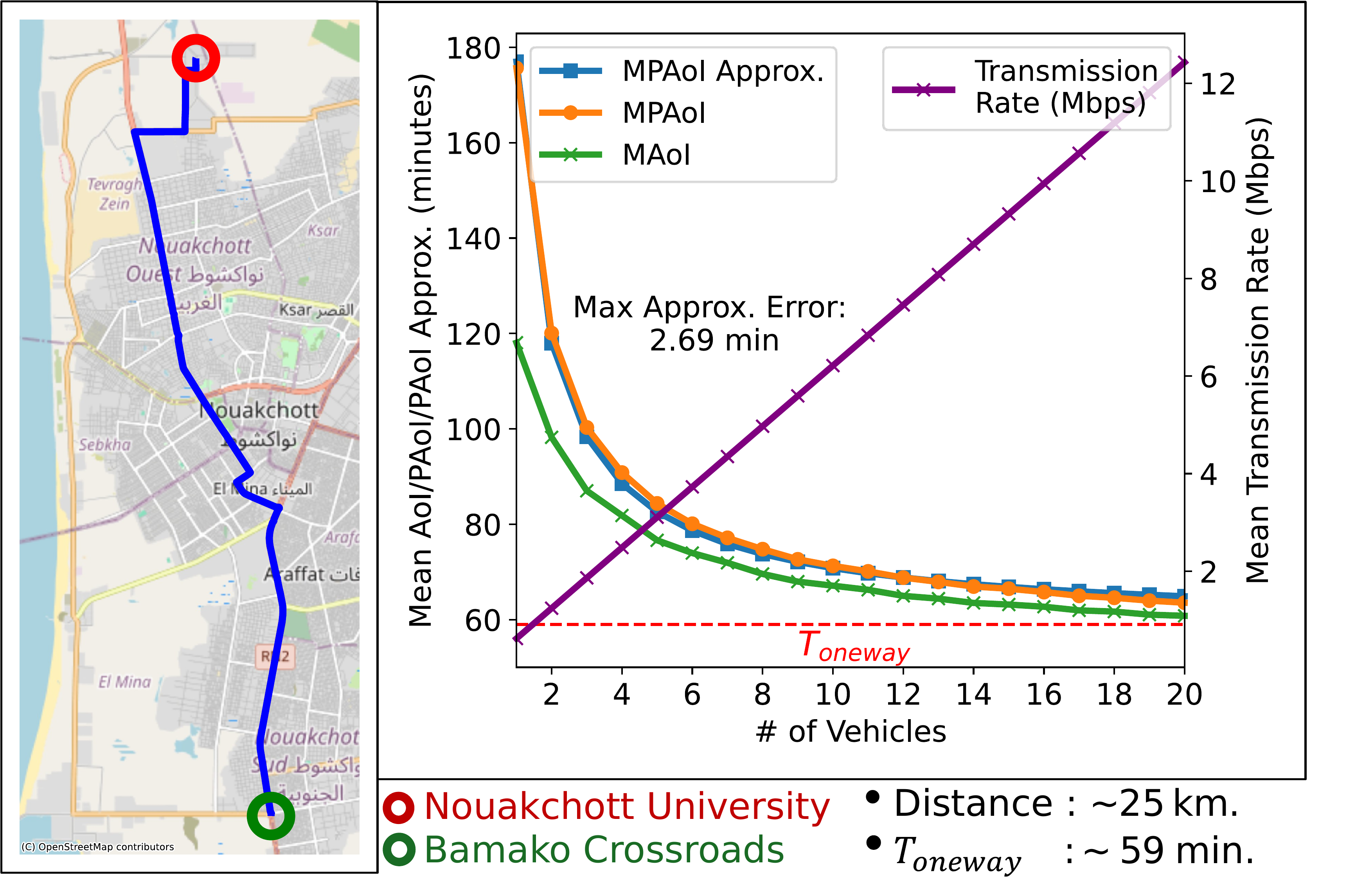}\label{fig:Nouakchott}}
	\subfigure[]
	{\includegraphics[width=1.0\linewidth]{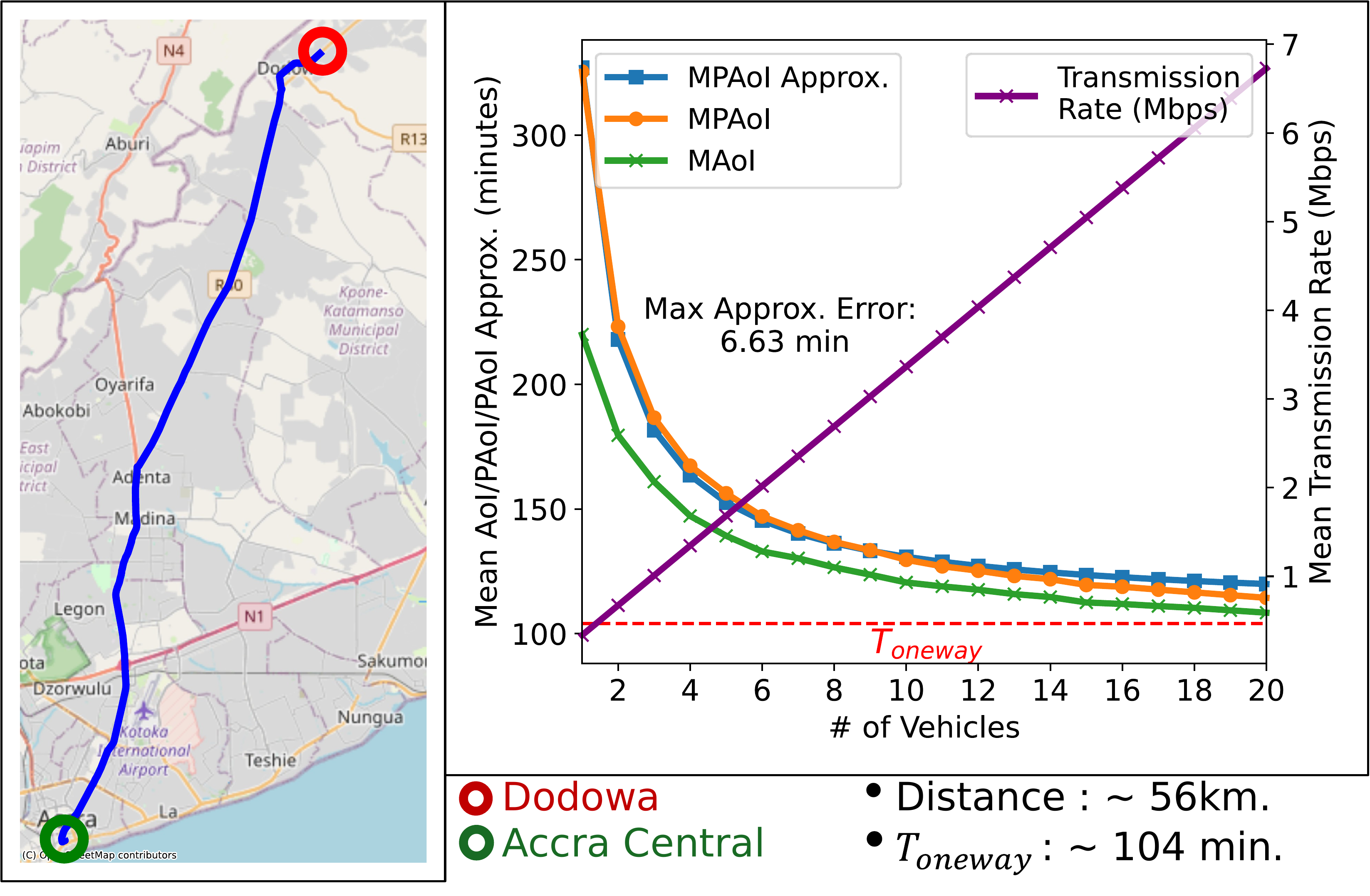}\label{fig:Accra}}
	\caption{Network performance metrics (Transmission Rate and MPAoI) for the routes (a) between \textit{Nouakchott University} and \textit{Bamako Crossroads}, and (b)  \textit{Dodowa} and \textit{Accra Central}, considering varying numbers of DTN data mules.}
	\label{fig:Nouakchott_Accra}
\end{figure}
\begin{figure}[h!]
    \centering
    \includegraphics[width=1.0\linewidth]{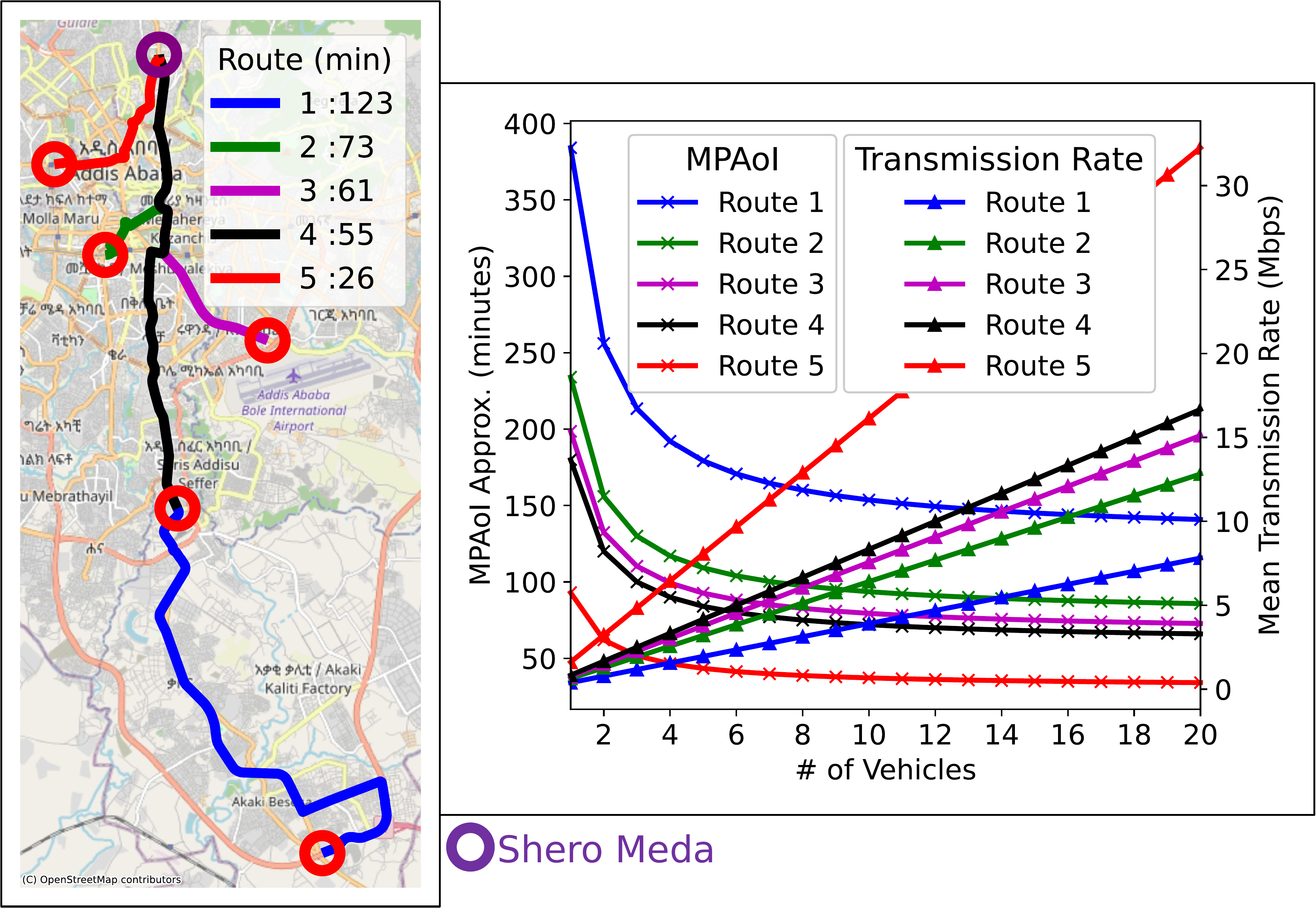}
    \caption{Performance evaluation of a star-shaped DTN with a central hub at \textit{Shero Meda}, connecting multiple routes in Addis Ababa with varying travel durations. The figure shows the impact of the number of DTN data mules on the data transmission rate and MPAoI for different routes, highlighting the interplay between network performance and route length.}
    \label{fig:Addis_Ababa}
\end{figure}

In Figs.~\ref{fig:Nouakchott_Accra} and \ref{fig:Addis_Ababa}, we utilize a recent dataset collected for informal public transport systems across 22 countries globally \cite{mittal2024efficient}. The dataset includes mobility traces for several routes, obtained by equipping vehicles with GPS transceivers to record the time required to complete their routes. While the dataset does not specifically provide information on vehicles starting and ending at bus stops or waiting times at the beginning and end of routes, it serves as a representative model of the randomness and delays commonly encountered in informal public transport systems when traveling between different locations. 
For this study, we selected the three longest routes in Nouakchott (capital of Mauritania), Accra (capital of Ghana), and Addis Ababa (capital of Ethiopia). Although some of the areas represented in our dataset have access to other connectivity solutions, we utilized this dataset as a means to demonstrate the effectiveness of our mule-based DTN framework for potentially bringing connectivity to rural areas.
We used the same contact time limits as listed in Table~\ref{tab:sim_params}, while the travel time statistics were derived from the dataset and the wireless link data rate is set to 20 Mbps. 
\par
Fig.~\ref{fig:Nouakchott} shows the results of our proposed framework applied to the route connecting Nouakchott University and \textit{Bamako Crossroads}. This route spans approximately 25 km and, under optimal conditions, takes around 30 minutes to complete a one-way trip. However, based on the dataset, informal transport vehicles take an average of 59 minutes to complete the one-way journey, accounting for random delays caused by various factors as analyzed earlier. The figure also illustrates the expected data transmission rate and MPAoI based on our framework. For a network with a single DTN data mule, the data transmission rate is around 1 Mbps, while it increases to approximately 12.5 Mbps with 20 data mules. Similarly, the MPAoI for one data mule is approximately 180 minutes, which decreases to around 65 minutes when 20 data mules are introduced into the network. These results highlight the interplay between the number of DTN data mules in the network and the resulting performance.
\par
A similar analysis is conducted for the route between \textit{Dodowa} and \textit{Accra Central} in \textit{Accra}, as shown in Fig.~\ref{fig:Accra}. This route is approximately 56 km long and requires around 55 minutes for a one-way trip under optimal conditions. However, based on the dataset, vehicles take an average of 104 minutes due to random delays. Because this route is longer, the network performance is slightly lower compared to the shorter route in \textit{Nouakchott} when using the same number of vehicles. For instance, with one DTN data mule, the transmission rate is less than 1 Mbps, and the MPAoI is approximately 350 minutes. With 20 DTN data mules, the data transmission rate improves to around 7 Mbps, while the MPAoI decreases to approximately 120 minutes. 

\par
Finally, we consider a scenario involving a central hub connecting multiple routes in Addis Ababa. This represents a central bus stop, \textit{Shero Meda}, serving as a hub for routes of varying lengths and travel durations. The hub connects routes to \textit{Tulu Dimtu} (Route 1: 123 minutes), \textit{Mexico} (Route 2: 73 minutes), \textit{Bole Millennium} (Route 3: 61 minutes), \textit{Abo Junction} (Route 4: 55 minutes), and \textit{Autobis Tera} (Route 5: 26 minutes), with the average one-way trip durations provided in parentheses. This can be modeled as a star-shaped DTN where the gateway at \textit{Shero Meda} serves multiple rural areas simultaneously.
The performance metrics for this setup are shown in Fig.~\ref{fig:Addis_Ababa}. For the route with the longest travel time (\textit{Tulu Dimtu}), the data transmission rate is less than 1~Mbps with a single DTN data mule, increasing to approximately 8~Mbps with 20 data mules. Similarly, the MPAoI decreases from 380~minutes with one data mule to around 150~minutes with 20 data mules. For shorter routes, consistent improvements in both transmission rate and MPAoI are observed as the number of data mules increases.

%%%%%%%%%%%%%%%%%%%%%%%%%%%%%%%%%%%%%%%%%%%%%%%%
%%%%%%%%%%%%%%%%%%%%%%%%%%%%%%%%%%%%%%%%%%%%%%%%
%%%%%%%%%%%%%%%%%%%%%%%%%%%%%%%%%%%%%%%%%%%%%%%%
\section{Conclusion}
%%%%%%%%%%%%%%%%%%%%%%%%%%%%%%%%%%%%%%%%%%%%%%%%
%%%%%%%%%%%%%%%%%%%%%%%%%%%%%%%%%%%%%%%%%%%%%%%%
%%%%%%%%%%%%%%%%%%%%%%%%%%%%%%%%%%%%%%%%%%%%%%%%
\par
This paper presented a DTN-based framework leveraging informal public transport systems to address the persistent digital divide in rural areas. This framework offers a cost-effective and scalable solution for enabling intermittent data exchange between urban regions with Internet access and rural regions without connectivity.
A probabilistic model was developed to characterize the movement of data mules, incorporating realistic travel and contact time distributions. Key performance metrics, including data transmission rates and Mean Peak Age of Information (MPAoI), were analyzed, and a mathematical approximation for the MPAoI was derived and validated through numerical simulations. The framework was further tested using real-world datasets from Nouakchott, Accra, and Addis Ababa, demonstrating its applicability in diverse settings.
The results quantify how the number of DTN-equipped vehicles and the route length influence achievable data transmission rates and data freshness, showing that additional data mules can improve performance even in resource-constrained settings.
\par
\textcolor{black}{While the investigated framework is well-suited for regions with limited infrastructure, its performance may decrease when vehicles travel very long distances between rural and urban areas, resulting in higher round-trip times and increased MPAoI. To address this limitation, future work could explore hybrid networking strategies that combine opportunistic DTN connections with infrastructure-based links, such as satellite backhaul, to maintain acceptable performance in remote or geographically isolated regions.}

\textcolor{black}{
%%%%%%%%%%%%%%%%%%%%%%%%%%%%%%%%%%%%%%%%%%%%%%%%
%%%%%%%%%%%%%%%%%%%%%%%%%%%%%%%%%%%%%%%%%%%%%%%%
%%%%%%%%%%%%%%%%%%%%%%%%%%%%%%%%%%%%%%%%%%%%%%%%
\section{Acknowledgement}
%%%%%%%%%%%%%%%%%%%%%%%%%%%%%%%%%%%%%%%%%%%%%%%%
%%%%%%%%%%%%%%%%%%%%%%%%%%%%%%%%%%%%%%%%%%%%%%%%
%%%%%%%%%%%%%%%%%%%%%%%%%%%%%%%%%%%%%%%%%%%%%%%%
The authors acknowledge support from the Arab Fund Programme at ICTP (ARF01 - AFESD Grant No. 14/2023) in the framework of the project ``Advancing the Capabilities of Arab Researchers and Students” and from the ICTP/IAEA Sandwich Training Educational Programme.
}

\appendices
\section*{Appendix} 
\subsection{Proof of \textit{Lemma}~\ref{lemma:mean_transmitted_data_size}} \label{section:Appendix_A}

Let \( X = \min(T_{c, A}, T_{c, B}) \), the CDF and PDF of the minimum of two i.i.d uniformly distributed random variables is:
\begin{align}
    &F_X(x) = P(X \leq x) = 1 - \left( \frac{c_2 - x}{c_2 - c_1} \right)^2, &\quad x \in [c_1, c_2]. \nonumber \\
    &f_X(x) = \frac{d}{dx} F_X(x)  = \frac{2(c_2 - x)}{(c_2 - c_1)^2}, &\quad x \in [c_1, c_2]. \nonumber
\end{align}
The mean of $X$ can be evaluated as: 
\[
E[X] = \int_{c_1}^{c_2} xf_X(x) \, dx =  \frac{2c_1+c_2}{3}.
\]
Hence, this completes the proof.

\ifCLASSOPTIONcaptionsoff
  \newpage
\fi

\bibliographystyle{IEEEtran}
\bibliography{references}

\begin{IEEEbiography}[{\includegraphics[width=1in,height=1.25in,clip,keepaspectratio]{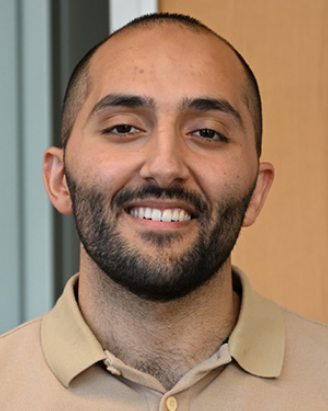}}]{Salah Abdeljabar}
(Graduate~Student~Member, IEEE) received the B.Sc. degree in electrical engineering from The University of Jordan, Amman, Jordan, in 2019, and the M.Sc. degree in electrical and computer engineering from the King Abdullah University of Science and Technology, Thuwal, Saudi Arabia, in 2023, where he is currently pursuing the Ph.D. degree. His research interests include long-range (LoRa) communication, Delay delay-tolerant networking (DTN), and optical wireless communications systems.
\end{IEEEbiography}

\begin{IEEEbiography}[{\includegraphics[width=1in,height=1.25in,clip,keepaspectratio]{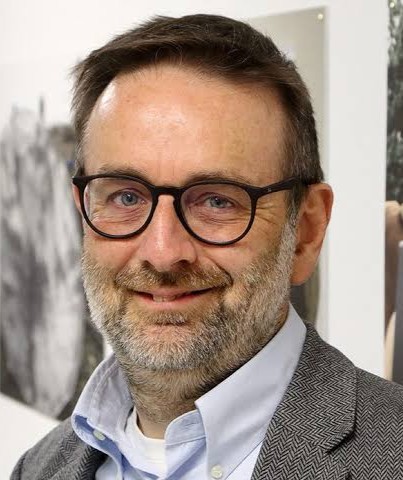}}]{Marco Zennaro}
(Senior~Member, IEEE) received the Ph.D. degree from
the KTH-Royal Institute of Technology, Stockholm, Sweden, and the M.Sc. degree in electronic engineering from the University of Trieste, Trieste, Italy.
He is a Researcher with the Abdus Salam International Centre for Theoretical Physics, Trieste, where he coordinates the Wireless Group of the Science, Technology and Innovation Unit.
He is a Visiting Professor with the
KIC-Kobe Institute of Computing, Kobe, Japan. His research interests include
ICT4D, the use of ICT for development, and in particular he investigates the use of IoT in developing countries. He has given lectures on wireless technologies in more than 30 countries.
\end{IEEEbiography}

\begin{IEEEbiography}[{\includegraphics[width=1in,height=1.25in,clip,keepaspectratio]{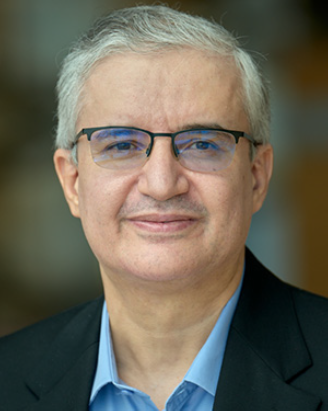}}]{Mohamed-Slim Alouini}
(Fellow, IEEE) was born in Tunis, Tunisia. He received the Ph.D. degree in electrical engineering from the California Institute of Technology, Pasadena, CA, USA, in 1998. He served as a Faculty Member with the University of Minnesota, Minneapolis, MN, USA, then with Texas A\&M University at Qatar, Doha, Qatar, before joining the King Abdullah University of Science and Technology, Thuwal, Makkah, Saudi Arabia, as a Professor of Electrical Engineering in 2009. His current research interests include the modeling, design, and performance analysis of wireless communication systems.
\end{IEEEbiography}

\end{document}